\documentclass[prd,twocolumn,floatfix,preprintnumbers,letterpaper,nofootinbib]{revtex4}
\usepackage{graphicx}
\usepackage{amsmath,amssymb}
\input{epsf}
\usepackage{epsf}
\usepackage{footnote}
\usepackage{natbib}

\begin{document}

\def\be{\begin{equation}}
\def\ee{\end{equation}}

\title{The Little Rip}
\author{Paul H. Frampton$^{1}$, Kevin J. Ludwick$^{1}$, and
Robert J. Scherrer,$^{2}$}
\affiliation{$^1$Department of Physics \& Astronomy, University of North
Carolina, Chapel Hill, NC~~27599}
\affiliation{$^2$Department of Physics \& Astronomy, Vanderbilt University,
Nashville, TN~~37235}
\date{\today}

\begin{abstract}
We examine models in which the dark energy density increases with time (so that
the equation-of-state parameter $w$ satisfies $w < -1$), but $w \rightarrow -1$
asymptotically, such that there is no future singularity.  We refine previous
calculations to determine the conditions necessary to produce this evolution.
Such models can display arbitrarily rapid expansion in the near future, leading
to the destruction of all bound structures (a ``little rip").  We determine
observational constraints on these models and calculate
the point at which the disintegration of bound structures occurs.  For the same
present-day value of $w$, a big rip with constant $w$ disintegrates bound structures
earlier than a little rip.

\end{abstract}

\maketitle

\section{Introduction}

Observations indicate 
that roughly 70\% of the energy density in the
universe is in the form of an exotic, negative-pressure component,
dubbed dark energy \cite{Knop,Riess}.  (See Ref. \cite{Copeland}
for a recent review.)  If $\rho_{DE}$ and $p_{DE}$ are the density
and pressure, respectively, of the dark energy, then
the dark energy can be characterized by the
equation-of-state parameter $w_{DE}$, defined by
\begin{equation}
w_{DE} = p_{DE}/\rho_{DE}.
\end{equation}

It was first noted
by Caldwell \cite{Caldwell} that
observational data do not rule out the possibility
that $w_{DE} <-1$.  Such ``phantom" dark energy
models have several peculiar properties.
The density of the dark energy {\it increases} with
increasing scale factor, and both the scale factor
and the phantom energy density can become infinite
at a finite $t$, a condition known as the ``big rip" \cite{Caldwell,rip,Taka,rip2}.
It has even been suggested that the finite lifetime for the universe
in these models may provide an explanation for
the apparent coincidence between the current values of the
matter density and the dark energy density \cite{doomsday}.

While $w(a) < -1$ as $a$ extends into the future is a necessary condition for a future singularity, it is
not sufficient.  In particular, if $w$ approaches $-1$ sufficiently rapidly,
then it is possible to have a model in which $\rho_{DE}$ increases with time,
but in which there is no future singularity.  Conditions which produce
such an evolution (specified in terms of $p_{DE}$ as a function
of $\rho_{DE}$) were explored in Refs. \cite{NOT,Stefancic}.

In this paper, we examine such models in more detail.  In particular, we will
extend the parameter space discussed in Refs. \cite{NOT,Stefancic} in both
directions, showing that there are nonsingular models in which $\rho_{DE}$ increases
more rapidly than the nonsingular models discussed in those references, and,
conversely, that there are singular models with $\rho_{DE}$ increasing
less rapidly than the singular models discussed in Refs. \cite{NOT,Stefancic}.
Models without a future singularity in which $\rho_{DE}$ increases with
time will nonetheless eventually lead to a dissolution of bound structures at some point in
the future, a process we have dubbed the ``little rip."  We discuss the
time scales over which this process occurs.  Finally, we consider observational
constraints on these models.

In the next section, we examine the conditions necessary for a future
singularity in models with $w < -1$.  In Secs. III and IV, specific little-rip models and 
disintegration of bound systems are studied.  Finally, in Sec. V, there is discussion.

\section{The Conditions for a Future Singularity}

We limit our discussion to a spatially flat universe, for which
the Friedmann equation is
\be
\label{Fried1}
\left(\frac{\dot a}{a}\right)^2 =  \frac{\rho}{3},
\ee
where $\rho$ is the total density,
$a$ is the scale factor, the dot will always
denote a time derivative,
and we take $\hbar = c = 8\pi G = 1$ throughout.
We will
examine the future evolution of our universe
from the point at which the pressure and density
are dominated by the dark energy, so we can
assume $\rho = \rho_{DE}$ and $p = p_{DE}$,
and for simplicity we will drop the $DE$ subscript.
Then the dark energy density evolves as
\be
\label{Fried2}
\dot{\rho} =  -3\left(\frac{\dot a}{a}\right)(\rho + p).
\ee

The simplest way to achieve $w < -1$ is to take a scalar
field Lagrangian with a negative kinetic term, and the conditions
necessary for a future singularity in such models
have been explored in some detail
\cite{Hao,Sami,Faraoni,Kujat}.  Here, however, we explore the more general
question of the conditions under which a dark energy density that increases
with time can avoid a future singularity, and the consequences of such
models.

One can explore this question from a variety of starting points, by specifying,
for example, the scale factor $a$ as a function of the time $t$ (an approach
taken, for example, in Refs. \cite{Barrow1,Barrow2,CV,BL}).
Alternately one can specify the pressure $p$
as a function of the density $\rho$, as in Refs. \cite{NOT,Stefancic}.
Note that this is equivalent to specifying the equation-of-state parameter
$w$ as a function
of $\rho$, since $w = p/\rho$.  Finally, one can specify the density $\rho$
as a function of the scale factor $a$.  Since we are interested specifically
in nonsingular models for which $\rho$ increases with $a$, we shall adopt
this last approach, but we will briefly examine the other two starting points.
Of course, given any one of these three functions, the other two can be derived
uniquely, but not always in a useful form.

For example, suppose that we specify $a(t)$.  In order to avoid a big rip,
it is sufficient that $a(t)$ simply be a nonsingular function for all $t$.
Writing
\begin{equation}
\label{a(t)}
a = e^{f(t)},
\end{equation}
where $f(t)$ is a nonsingular function,
the density is given by equation (\ref{Fried1}) as
$\rho = 3(\dot a/a)^2 = 3\dot f^2$, and the condition that $\rho$ be an increasing
function of $a$ is simply $d \rho/da = (6/\dot a) \dot f \ddot f > 0$,
which is satisfied as long as
\begin{equation}
\label{ddf}
\ddot f > 0.
\end{equation}
Thus, all little-rip models are
described by an equation of the form (\ref{a(t)}), with nonsingular $f$
satisfying equation (\ref{ddf}).

Now consider the approach of Refs. \cite{NOT,Stefancic}, who expressed the pressure
as a function of the density in the form
\begin{equation}
p = - \rho - f(\rho),
\end{equation} 
where $f(\rho) > 0$ ensures that the $\rho$ increases with scale factor.  
In order to determine the existence of a future singularity, one can integrate
equation (\ref{Fried2}) to obtain \cite{NOT}
\begin{equation}
a = a_0 \exp\left(\int \frac{d\rho}{3f(\rho)}\right),
\end{equation}
and equation (\ref{Fried1}) then gives \cite{NOT}
\begin{equation}
\label{trho}
t = \int \frac{d\rho}{\sqrt{3 \rho} f(\rho)}.
\end{equation}
The condition for a big-rip singularity is that the integral
in equation (\ref{trho}) converges.  Taking a power law for $f(\rho)$, namely
\begin{equation}
f(\rho) = A \rho^{\alpha},
\end{equation}
we see that a future singularity can be avoided for $\alpha \le 1/2$
\cite{NOT,Stefancic}.  We examine this boundary in more detail below,
noting that one can have $f(\rho)$ increase more rapidly than $\rho^{1/2}$
without a future singularity.

Now consider the third possibility:  specifying the density
$\rho$ as an increasing function of scale factor $a$.  We will seek upper and lower bounds on the growth
rate of $\rho(a)$ that can be used to determine whether or not a big-rip singularity is
produced.  Defining $x \equiv \ln a$, we can rewrite equation (\ref{Fried1}) as
\begin{equation}
t = \int \sqrt\frac{3}{\rho(x)} dx,
\end{equation}
and the condition for avoiding a future big-rip singularity is
\begin{equation}
\label{condition}
\int_{x_0}^\infty \frac{1}{\sqrt{\rho(x)}} dx \rightarrow \infty.
\end{equation}
The case $p = - \rho - A\rho^{1/2}$ from Refs.
\cite{NOT,Stefancic} corresponds to 

\begin{equation}
\label{model1}
\frac{\rho}{\rho_0} =  \left(\frac{3A}{2\sqrt {\rho_0}} \ln (a/a_0) + 1 \right)^2,
\end{equation}
where $w \le -1$ requires $A \geq 0$, and we take $\rho = \rho_0$ and $a = a_0$ at a fixed time $t_0$.
Expressing this density as a function of time rather than scale factor
gives a much simpler expression:
\begin{equation}
\frac{\rho}{\rho_0}  = e^{\sqrt{3}A(t-t_0)}.
\end{equation}
The equation-of-state parameter $w$ corresponding to equation (\ref{model1})
can be derived from the relation $(a/\rho)(d\rho/da) = -3(1+w)$:
\begin{equation}
w = -1 - \frac{1}{\frac{3}{2}\ln(\frac{a}{a_0}) + \frac{\sqrt{\rho_0}}{A}},
\end{equation}
and the corresponding
expansion law is
\begin{equation}
\frac{a}{a_0} = e^{(2\sqrt{\rho_0}/3A)[e^{(\sqrt{3}A/2)(t-t_0)}-1]}.
\label{AAA}
\end{equation}

However, we can find $\rho(a)$ for which $\rho$ increases more rapidly
with $a$, but for which equation (\ref{condition}) is still satisfied.
For example, writing $\rho^{1/2} \sim (\ln a)(\ln \ln a)$ as $a \rightarrow
\infty$ satisfies
equation (\ref{condition}).  An example of such a $\rho$, with a free
parameter B, is 
\begin{equation}
\label{model2}
\frac{\rho}{\rho_0} = 
N\left(\frac{a}{a_0},B\right) 
\frac{(1+\ln(\frac{a}{a_0}+B))^2}{(1+\ln(1+B))^2}
\frac{(\ln(1+ \ln(\frac{a}{a_0}+B)))^2}{(\ln(1+ \ln(1+B)))^2},
\end{equation}
where the choice
\begin{equation}
N\left(\frac{a}{a_0}, B\right) = \frac{(\frac{a}{a_0}+B)^2}{(1+B)^2 (\frac{a}{a_0})^2}
\label{N}
\end{equation}
leads to a real, nonnegative $\rho$ and an analytic form for the behavior of $a(t)$:
\begin{equation}
\frac{a}{a_0} =  e^{( e^{ { \ln(1+\ln(1+B)) e^{\left[ \frac{\sqrt{\rho_0/3}(t-t_0)}{(1+B)(1+\ln(1+B)) 
\ln(1+\ln(1+B))} \right]}}} -1)} -B.
\label{BBB}
\end{equation}

This argument can be extended further.
In general, if we denote
$\ln_j (x) \equiv \ln \ln \ln....\ln (x)$, where the logarithm on the right-hand side
is iterated $j$ times, then any function of the form
\begin{equation}
\label{nestedlog}
\rho \sim (\ln a)^2(\ln_2 a)^2(\ln_3 a)^2...(\ln_m a)^2
\end{equation}
satisfies equation (\ref{condition}) as $a \rightarrow \infty$ 
and avoids a big-rip singularity.
A density increasing as in equation (\ref{nestedlog}) leads to an expansion
law of the form
\begin{equation}
\label{nestedexp}
a \sim \exp( \exp( \exp...(\exp(t))...)),
\end{equation}
where there are $m+1$ exponentials.  We have omitted the constants in equations
(\ref{nestedlog}) and (\ref{nestedexp}) for the sake of clarity.
Equation (\ref{nestedexp}), while growing extraordinarily rapidly, is manifestly
nonsingular.  While an expansion law of this sort might seem absurd, it is
probably less so than a big-rip expansion law, and in any case our goal
is to try to determine the boundary between little-rip and big-rip evolution for
$\rho(a)$.  In this spirit, consider the slowest growing power-law modification
to equation (\ref{nestedlog}):
\begin{equation}
\label{nestedrip}
\rho \sim (\ln a)^2(\ln_2 a)^2(\ln_3 a)^2...(\ln_m a)^{2+\epsilon},
\end{equation}
where $\epsilon > 0$ is a constant.  No matter how small $\epsilon$ is,
and despite the fact that it modifies an extraordinarily slowly growing
nested logarithm function, the growth law in equation (\ref{nestedrip}) leads to a future
big-rip singularity.

Note that the bounds specified by equations (\ref{nestedlog}) and
(\ref{nestedrip}) are not sharp; we can always find forms for $\rho(a)$ that
interpolate between these two behaviors and produce either a little
rip or a big rip.
However, as we take $m$ to be arbitrarily large, nearly any function of
interest will increase more rapidly than equation (\ref{nestedlog}) or
more slowly than equation (\ref{nestedrip}), allowing us a practical, if
not a rigorously sharp, bound.  This lack of a sharp bound is 
due to the fact that there is
no bound on the fastest growing function $a(t)$ which is nonsingular at finite $t$.

If one is willing to place other restrictions on the form of $\rho(a)$, then
more stringent bounds apply.  Barrow \cite {Barrow3} demonstrated that if
$\rho + 3 p$ is a rational function of $a$ and $t$, and $a(t)$ is
nonsingular at finite $t$, then $a(t)$
can grow no more rapidly than the double exponential of a polynomial in $t$.
Our equation (\ref{nestedlog}) violates this condition because of the
logarithmic functions.

\section{Constraining little-rip models}

Here we shall examine in more detail the two specific little-rip models
given by equations (\ref{model1}) and (\ref{model2}), which we will
call model 1 and model 2, respectively.
Note that we do {\it not} make use of equations (\ref{AAA}) and (\ref{BBB})
here,
as these are valid only when the matter density can be neglected in comparison
to the dark energy density.
Model 1 is characterized by a single free parameter $A$,
and the scale factor behaves asymptotically as a double exponential in $t$,
as in equation (\ref{AAA}):
\begin{equation}
a ( t)  \stackrel{t\rightarrow +\infty}{\longrightarrow} e^{e^t}
\label{modelA}
\end{equation}
The parameter $A$ is chosen to make a best fit
to the latest supernova data from the Supernova Cosmology Project \cite{SCP}, and
has the best-fit value $A = 3.46 \times 10^{-3} {\rm Gyr}^{-1}$,
while a $95\%$ C.L. fit 
can be found for the range $-2.74 \times 10^{-3} {\rm Gyr}^{-1} \leq A \leq 
9.67 \times 10^{-3} {\rm Gyr}^{-1}$. 

Model 2 is characterized by the free parameter $B$ and
has a scale factor that behaves asymptotically as a triple exponential in $t$,
as in equation (\ref{BBB}): 
\begin{equation}
a ( t)  \stackrel{t\rightarrow +\infty}{\longrightarrow}  e^{ e^{e^t}}
\label{modelB}
\end{equation}
The parameter  $B$ is chosen to make a best fit
to \cite{SCP} as well, and it
has the value $B = 1.23$.  The confidence interval for $B$ at the $95\%$ C.L. 
is $1.12 \leq B \leq 1.34$.  In fitting both models, $\Omega_{m_0} = 0.274$, 
$\Omega_{x_0} = 1- \Omega_{m_0}$, and $H_0 = 70.1$ km s$^{-1}$ Mpc$^{-1}$, which are 
consistent with the best-fit ranges for these values given by WMAP \cite{WMAP}.
The resultant
Hubble and residual $\Lambda$CDM ($w = -1$) plots of distance
modulus $\mu$ versus redshift $z$ for both models are displayed in Fig. 1.

\begin{figure}[htp]
\begin{center}
\fbox{
{\label{Hubble}\includegraphics[scale=0.60]{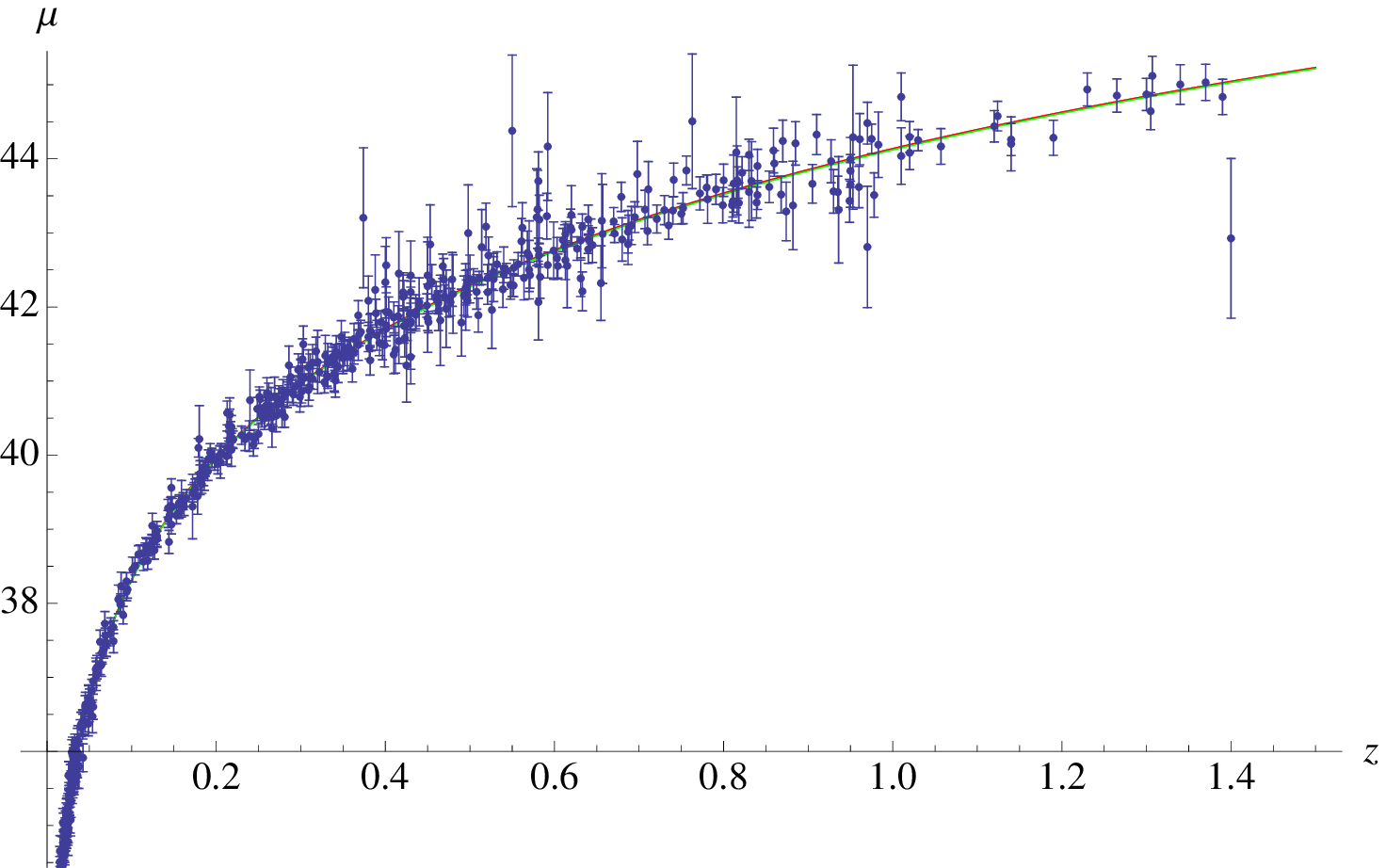}}} \\
\fbox{
{\label{residual}\includegraphics[scale=0.60]{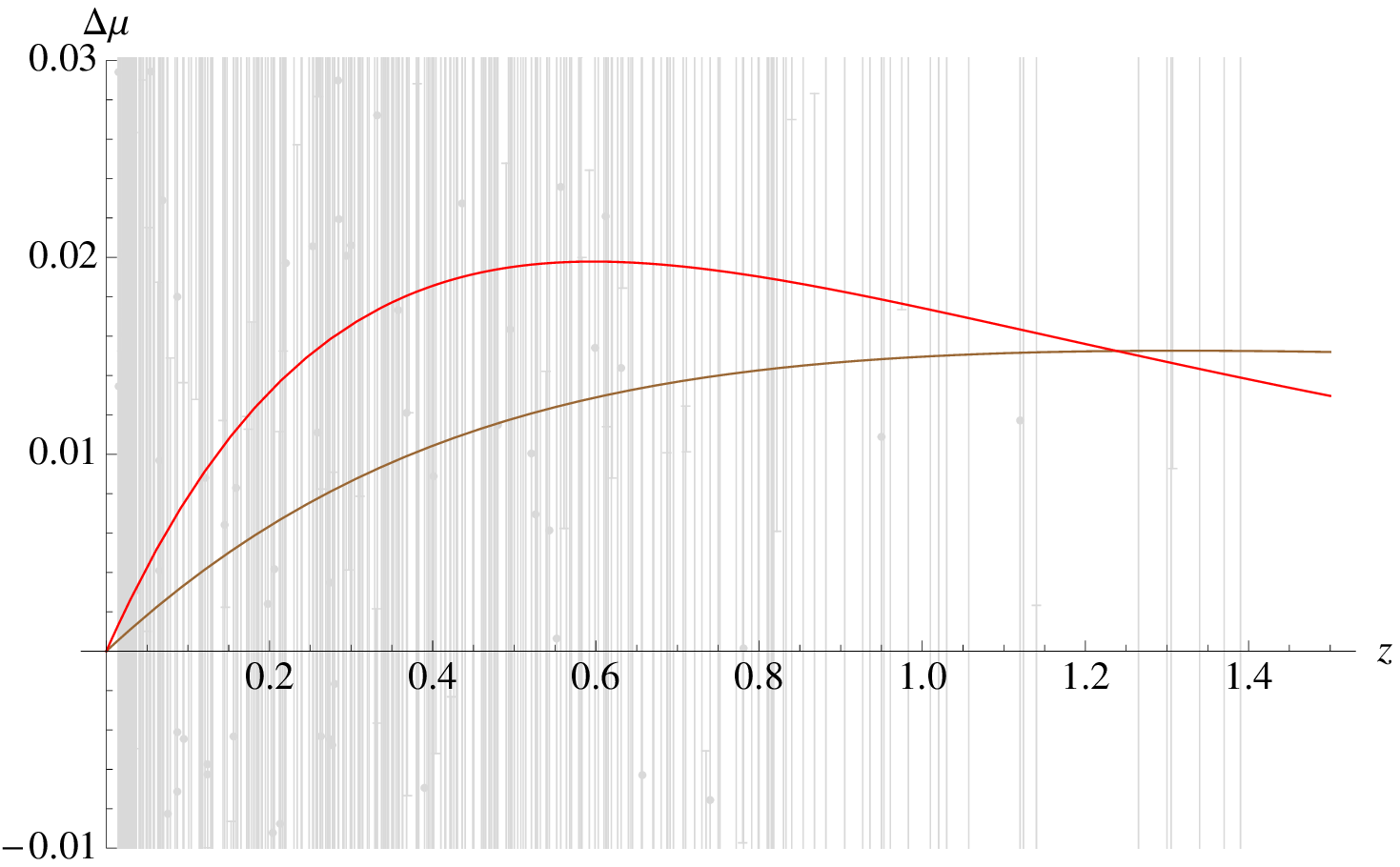}}}
\caption{Top panel:  Hubble plot of distance modulus $\mu$ versus redshift $z$ for the
$\Lambda$CDM ($w = -1$) model (green) and models 1 (brown) and 2
(red).  The lines are essentially 
indistinguishable.  Bottom panel:  The $\Lambda$CDM model is 
subtracted from models 1 (brown) and 2 (red).  The $\Lambda$CDM model is, by 
definition, represented by the 
$\Delta \mu = 0$ axis.  As can be judged by the 
size of the error bars of the data, 
all are excellent fits to the supernovae data.  (The color plots 
are in the online version of the paper.)}
 
\label{fig1}
\end{center}
\end{figure}

Not surprisingly, the best-fit models closely resemble 
the $\Lambda$CDM
model, which is known to be an excellent fit to the data \cite{Melchiorri}.  To see this more
clearly, note that our models will resemble a cosmological constant
at low redshift as long as $\rho(a) \sim constant$ for $a \sim a_0$.  For
model 1, this condition is satisfied when $A/\sqrt{\rho_0} \ll 1$ in equation (\ref{model1}),
while for model 2, we require $B \simeq 1.39$ in equation
(\ref{model2}).  To see that $B$ should be close to this value, 
one should expand equation (\ref{model2}) around $a = a_0$.  The 
zeroth-order term is $\rho_0$, and the coefficient for the first-order 
term is $0$ when $B = 1.39$.  A comparison with our best-fit values indicates that these
conditions are, indeed, satisfied.  Furthermore, in the limit where these
conditions are satisfied, these little-rip models closely resemble, at low
redshift,
big-rip models close to $\Lambda$CDM, i.e., models with constant $w < -1$ and
$|1+w| \ll 1$.
To see this,
recall that constant-$w$ big-rip models have a density varying with
scale factor as
\begin{equation}
\label{rhorip}
\rho = \rho_0 (a/a_0)^{-3(1+w)}.
\end{equation} 
For $|1+ w| \ll 1$ and $a/a_0$ not too far from 1, equation
(\ref{rhorip}) behaves as
\begin{equation}
\label{rhorhip2}
\rho \approx \rho_0 [1-3(1+w) \ln(a/a_0)].
\end{equation}
Equation (\ref{model1}) reduces to equation (\ref{rhorhip2})
for $A/\sqrt{\rho_0} \ll 1$, with
$A/\sqrt{\rho_0} = -(1+w)$.

\section{Disintegration}

A feature of a big rip is that all bound-state systems disintegrate
before the final singularity \cite{rip}. Here we show that
little-rip models, despite not having a final singularity, also produce
the disintegration of bound structures.
As a first approximation, the disintegration
time is when the dark energy density equals the mean density of the system.
A more accurate method was presented in \cite{rip2}. We shall employ both
methods to estimate the disintegration of the Sun-Earth system.\footnote{When
the Sun becomes a red giant in $\sim 5$ Gyrs, it will envelope Mercury
and Venus, and (maybe) Earth \cite{sun}. Here, for the sake of making a point,
we assume the Earth will continue to orbit the Sun until unbound by dark energy.}
For the little-rip models 1 and 2, with the best-fit parameters
derived in the previous section, we find the time $t_{\odot-\oplus}$ from
the present time $t_0$ until the Earth ($\oplus$) - Sun ($\odot$) system is disintegrated
to be:
\begin{equation}
{\bf {\rm Model~ 1:}} ~~~~~~ t_{\odot-\oplus} \simeq 8~ {\rm Tyrs} 
\label{LRA}
\end{equation}
\begin{equation}
{\bf {\rm Model~ 2:}} ~~~~~~ t_{\odot-\oplus} \simeq 146~ {\rm Gyrs}. 
\label{LRB}
\end{equation}
\noindent
Note that the disintegration time for model 2 is less than 
that of model 1, which is expected since $\rho$ for model 2
grows faster than $\rho$ for model 1.  

It is straightforward to estimate the corresponding $t_{\odot-\oplus}$ for
big-rip models with constant $w$ to be \cite{Taka}
\begin{equation}
t_{\odot-\oplus} \simeq \left( \frac{11~ {\rm Gyrs}}{|1 + w|} \right),
\label{BR}
\end{equation}
and it is almost identical to $t_{rip}$, which is about one year later.

Clearly, little-rip models can produce this disintegration either earlier or
later than big-rip models, depending on the exact parameters of each model.
For example, by putting, $w = -1 - 10^{-3}$ in equation (\ref{BR}), we
find a value of $11$ Tyrs for $t_{\odot-\oplus}$, which is larger than that of models 1 and 2 
in Eqs.(\ref{LRA}, \ref{LRB}). In this case, disintegration occurs
earlier in the little-rip model than in the
big-rip model.

The five energy conditions (weak, null, dominant, null dominant, strong) 
(see, e.g., Ref. \cite{Carroll})  are all violated
by all little-rip and big-rip models. A simple way to see this is that if
$w < -1$, which occurs for any rip, a boost is allowed
with $(v/c)^2 > - w/c$ to an inertial frame with negative energy density.
Having said that, if general relativity itself fails for length scales bigger
than that of galaxies, we may not be constrained by the same energy conditions.

\section{Discussion}

In the big rip, the scale factor and density diverge in a singularity at a finite future
time. In the $\Lambda$CDM model,
there is no such divergence and no disintegration because the dark energy
density remains constant. The little rip interpolates
between these two cases; mathematically it can be represented as an 
infinite limit sequence which
has the big rip and the $\Lambda$CDM model as its boundaries.  Such models
can be represented generically by a density varying with scale factor as in
equation (\ref{nestedlog}).

Physically, in the
little rip, the scale factor and the density are never infinite at a finite
time. Nevertheless, such models generically lead to structure disintegration at a finite
time.  For models consistent with current supernova observations, such disintegration
can occur either earlier or later in a little-rip model than in a big-rip model, depending
on the parameters chosen for the models.
However, for a given present-day value of $w$, the big-rip model
with constant $w$ will necessarily lead to an earlier disintegration than the
little-rip
model with the same present-day value of $w$.  This results from the fact that
$w$ increases monotonically in the little
rip models, resulting in a smaller value for $\rho$ at any given $a$ than in the corresponding
constant-$w$ big-rip model, and
therefore, a lower expansion rate.  Thus, supernova bounds on the epoch of
disintegration for constant-$w$ big-rip models also apply to little-rip models; one cannot
simultaneously
satisfy supernova constraints and hasten the onset of disintegration to an arbitrarily early time
simply by iterating exponentials in the expansion law.

Furthermore, supernova data
force both big-rip and little-rip models into a region of parameter space in which
both models resemble $\Lambda$CDM.  In this limit, big-rip and little-rip models
produce essentially the same expansion law
up to the present, despite having very different future evolution.
Thus, current data already make it essentially impossible to determine
whether or not the universe will end in a future singularity.

Finally, we remark that since the novel and speculative cyclic
cosmology proposed in Ref. \cite{BF} requires only disintegration and
not a singularity,
such cyclicity would seem to be possible within a little-rip model
instead of the big rip considered in \cite{BF}. This is one
potentially fruitful direction for future research \cite{FLS2}.

\acknowledgements 

We thank Ryan M. Rohm for useful discussions.  P.H.F. and K.J.L. were 
supported in part by the Department of Energy (DE-FG02-05ER41418).
R.J.S. was supported in part by the Department of Energy (DE-FG05-85ER40226).

\end{document}